\documentclass[aip,apl,reprint]{revtex4-1}
\pdfoutput=1
\usepackage[version=3]{mhchem} 
\usepackage{epsfig}
\usepackage{graphicx} 

\usepackage{dcolumn} 
\usepackage{bm} 
\usepackage{amssymb} 
\usepackage{amsmath}

\usepackage{siunitx}

\usepackage{url}
\usepackage{color}

\usepackage{epstopdf}
\begin{document}

\author{David P. Lake}
\affiliation{Institute for Quantum Science and Technology, University of Calgary, Calgary, AB, T2N 1N3, Canada}
\author{Matthew Mitchell}
\affiliation{Institute for Quantum Science and Technology, University of Calgary, Calgary, AB, T2N 1N3, Canada}
\affiliation{National Institute for Nanotechnology, 11421 Saskatchewan Dr.\ NW, Edmonton, AB T6G 2M9, Canada}
\author{Harishankar Jayakumar}
\affiliation{Department of Physics, City College of New York, New York, NY 10031, USA}
\author{La\'is Fujii dos Santos}
\affiliation{Instituto de F\'isica Gleb Wataghin, Universidade Estadual de Campinas, 13083-970 Campinas, S\~ao Paulo, Brazil}
\author{Davor Curic}
\affiliation{Institute for Quantum Science and Technology, University of Calgary, Calgary, AB, T2N 1N3, Canada}
\author{Paul E. Barclay}
\email{pbarclay@ucalgary.ca}
\affiliation{Institute for Quantum Science and Technology, University of Calgary, Calgary, AB, T2N 1N3, Canada}
\affiliation{National Institute for Nanotechnology, 11421 Saskatchewan Dr.\ NW, Edmonton, AB T6G 2M9, Canada}

\title{Efficient telecom to visible wavelength conversion in doubly resonant GaP microdisks}

\begin{abstract}
Resonant second harmonic generation between 1550 nm and 775 nm with normalized outside efficiency $> 3.8 \times10^{-4} \,\text{mW}^{-1}$ is demonstrated in a gallium phosphide microdisk  supporting high-$Q$ modes at visible ($Q \sim 10^4$) and infrared ($Q \sim 10^5$) wavelengths. The double resonance condition is satisfied for a specific pump power through intracavity photothermal temperature tuning using $\sim 360\,\mu$W of 1550 nm light input to a fiber taper and coupled to a microdisk resonance. Power dependent efficiency consistent with a simple model for thermal tuning of the double resonance condition is observed.
\end{abstract}

\maketitle

Since the first observation of second harmonic generation (SHG) in 1961 \cite{ref:Franken}, it has become a ubiquitous demonstration of nonlinear optics. In recent years there has been mounting interest in SHG within micron-scale optical structures such as waveguides and cavities \cite{ref:nakagawa1995shg,ref:mondia2003esh,ref:ilchenko2004noc, ref:McCutcheon2007,ref:Yamada2014, ref:Mariani:2013, ref:Buckely2013, ref:Diziain2013, ref:Kuo2011,ref:Wang2014, ref:Kuo2014}. These experiments seek to take advantage of devices whose combination of large optical quality factor, $Q$, and small mode volume, $V$, provide enhancements to electromagnetic per-photon field intensities at both fundamental and second harmonic wavelengths. Furthermore, the compact nature of these optical devices lends itself to convenient integration into complex on-chip photonic circuits. To date SHG has been demonstrated in a number of microresonator geometries including microdisks \cite{ref:Mariani:2013,ref:Kuo2014,ref:Wang2014,ref:Mariani2014}, microrings \cite{ref:Pernice2012}, microspheres \cite{ref:Dominguez-Juarez2011}, photonic crystal nanocavities \cite{ref:McCutcheon2007,ref:Yamada2014,ref:Buckely2013,ref:Diziain2013} and waveguides \cite{ref:Rivoire2011,ref:lengle2013}. An impressive $9 \times 10^{-2} \text{mW}^{-1}$ conversion efficiency was achieved by Fu\"rst et.\ al.\ in macroscopic (mm) sized whispering gallery mode resonators \cite{ref:Furst2010}. Conversion between $1985 \ \text{nm}$ and $993 \ \text{nm}$ light with normalized outside and circulating efficiencies of $10^{-5}\text{mW}^{-1}$ and $10^{-3} \text{mW}^{-1}$, respectively, have been demonstrated in $\mu\text{m}$-scale gallium arsenide microdisks with low optical absorption at IR wavelengths \cite{ref:Kuo2014}. Microcavities have also been utilized in nonlinear optical processes including sum-frequency generation \cite{ref:Buckley2014}, and third harmonic generation \cite{ref:Carmon2007}.

\begin{figure}[t]
\begin{center}
\epsfig{figure=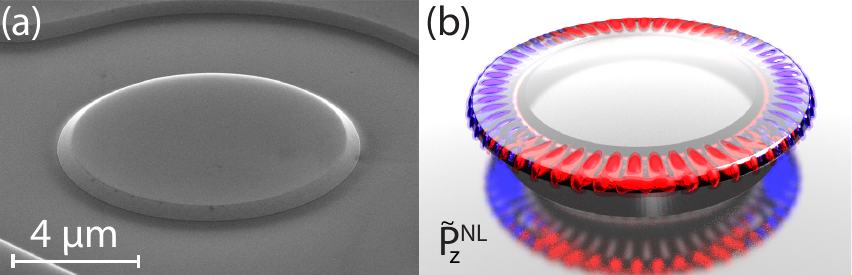, width=1\linewidth}
\caption{(a) SEM of a GaP microdisk before being undercut. (b) $\widetilde{\mathsf{P}}_z^{\mathsf{NL}}$ generated by the $m = 27$ fundamental TE mode at $\lambda_c \sim 1550\,\text{nm}$ of a microdisk with dimension as in (a).}
\label{fig:modes}
\end{center}
\end{figure}

In this work we study microdisk cavities such as the device shown in Fig.\ \ref{fig:modes}{(a)}, fabricated from gallium phosphide (GaP). The optical transparency window of GaP spans wavelengths from 550\,nm to IR, making it a promising material for nonlinear wavelength conversion between C-band (1550 nm) and visible wavelengths \cite{ref:Buckely2013}. Second harmonic conversion efficiency can be enhanced by fabricating microcavities supporting high-$Q$ optical resonances at $\lambda_c$ and $\widetilde\lambda_c$ that are resonant with the pump ($\lambda_o$) and second harmonic wavelengths ($\widetilde{\lambda}_o=\lambda_o/2$), respectively. The high sensitivity of small-$V$ devices to fabrication imperfections and variations in material optical properties, combined with their large free-spectral range, makes it challenging in practice to realize high-$Q/V$ doubly-resonant microcavities. Approaches for addressing this problem include near-field perturbative tuning \cite{ref:Shambat2010}, and bulk temperature tuning \cite{ref:Mariani2014}. Here we demonstrate a method for tuning optical resonance wavelengths in-situ using dispersive intracavity thermo-optic effects, and utilize this tuning to enable highly-efficient conversion between $\lambda_o = 1545\,\text{nm}$ and $\widetilde{\lambda}_o = 772\,\text{nm}$, with normalized outside efficiency in excess of $3.8\times10^{-4}$ \SI{}{mW}$^{-1}$.

Efficient SHG in microcavities requires, in addition to resonances at $\lambda_o$ and $\widetilde{\lambda}_o$, a phasematching mechanism to overcome dispersion intrinsic to the microcavity. A common approach to achieve this, known as quasi-phasematching, relies upon periodic domains with alternating nonlinear susceptibility \cite{ref:boyd2003no}. The zincblende structure of GaP possess $\overline{4}$ symmetry \cite{ref:boyd2003no}, which can be used to realize quasi-phasematching of microdisk whispering gallery modes without explicit creation of periodic domains \cite{ref:dumeige2006wgm, ref:Kuo2011, ref:Kuo2014}. The microdisks studied here support radially polarized (TE) modes near ${\lambda}_o$, and both TE and vertically ($\hat{z}$) polarized (TM) modes near $\widetilde{\lambda}_o$, that are coupled by the second order nonlinear susceptibility of GaP. The  amplitude of the radial and $\hat{z}$ field components these modes varies with $e^{im\theta}$, where $m$ is the azimuthal mode number and $\theta$ is the cylindrical angular coordinate. The corresponding cartesian in-plane components $E_x$ and $E_y$ of the TE modes experiences a sign change with period $|\Delta\theta| = \pi$. As a result, the second order nonlinear polarization along $z$, $\widetilde{\mathsf{P}}_z^{\mathsf{NL}} = 2\epsilon_0 d_{14} E_x E_y$ where $d_{14}$ is the relevant nonlinear susceptibility tensor element of GaP, experiences a sign change with period $|\Delta\theta| = \pi/2$. This is visualized in Fig. \ref{fig:modes}(b), and can be interpreted as momentum imparted by the periodic effective inversion of crystal orientation, creating azimuthal momentum components at $2m\pm2$. As a result, the time-averaged coupling between $\widetilde{\mathsf{P}}_z^{\mathsf{NL}}$ and the  TM microdisk mode with azimuthal index $\widetilde{m}$ is maximized when $\widetilde{m}=2m\pm2$. This $\pm2$ offset can compensate for microdisk dispersion\cite{ref:Kuo2011}, so that satisfying  $\lambda_c|_{m} = 2\widetilde\lambda_c|_{2m\pm2}$ becomes possible.

The impact of these effects on SHG is captured by \cite{ref:Kuo2011,ref:lake2015sup}:
\begin{align}\label{eq:eff}
\widetilde{P} =   |K|^2 & \frac{\widetilde{\lambda}_c/2\widetilde{Q}_e}{\widetilde\Delta(\lambda_o, P_d)^2+(\widetilde{\lambda}_c/2\widetilde{Q}_t)^2} \\ & 
\left[  \frac{\lambda_c/2Q_e}{\Delta(\lambda_o, P_d)^2+\lambda_c^2/4Q_t^2} \right]^2 P^2, \nonumber
\end{align}
which describes the second harmonic power $\widetilde{P}$ output into a waveguide coupled to the microdisk for IR pump power $P$ input to the waveguide.  $Q_{e,t}$ and $\widetilde{Q}_{e,t}$ are the external waveguide coupling ($e$) and the total ($t$) quality factors of the pump and IR microdisk modes, respectively. The quasi-phasematching is captured by the second harmonic coefficient, $K$, given in the Supplementary Material \cite{ref:lake2015sup}.  The remaining terms describe the cavity enhancement, and are maximized at ``double resonance'' where both the pump and second harmonic fields are resonant with a cavity mode, i.e.\    $\Delta = \widetilde\Delta = 0$. Here, $\Delta(\lambda_o, P_d) ={\lambda_o - \lambda_c(P_d)}$ and $\widetilde\Delta(\lambda_o, P_d) ={\widetilde\lambda_o -\widetilde\lambda_c(P_d)} = \lambda_o/2 -\widetilde\lambda_c(P_d)$ are the detunings between the pump and second harmonic fields, respectively, and the microdisk modes. Due to photothermal effects discussed below, $\lambda_c$ and $\widetilde\lambda_c$ depend on the pump power dropped into the cavity, $P_d$.

\begin{figure}[t]
\begin{center}
\epsfig{figure=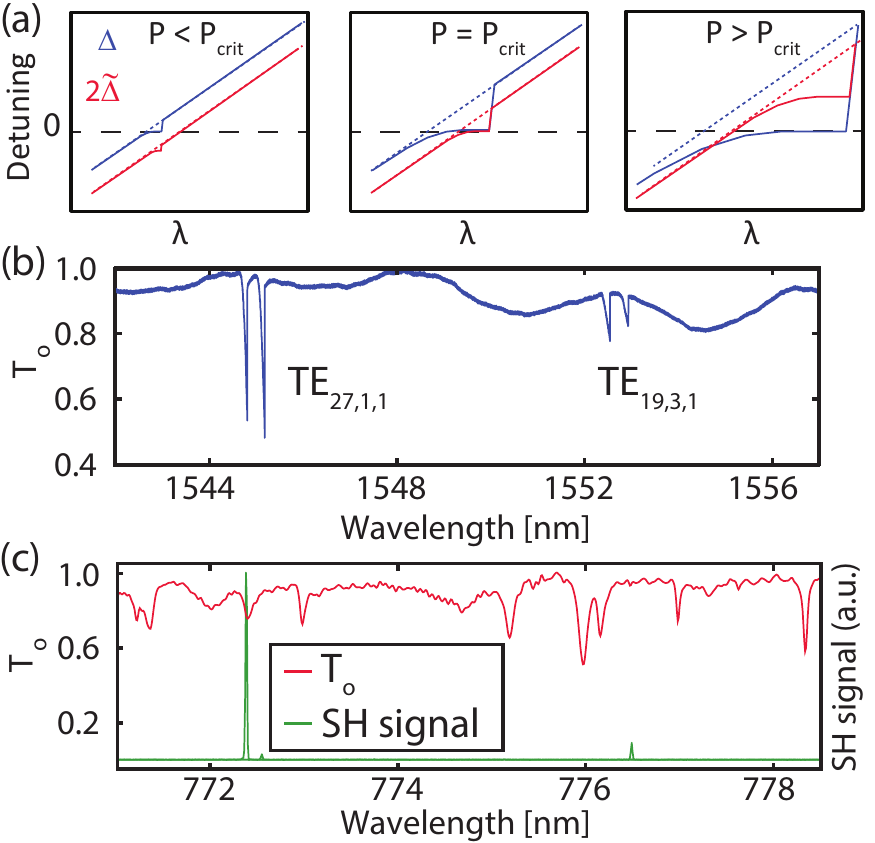, width=1\linewidth}
\caption{(a) Cartoon of influence of $P$ on detunings $\Delta(\lambda_o; P)$ (blue) and $2\widetilde\Delta(\lambda_o; P)$ (red) for $\eta > 0$.  The dashed lines show the expected behavior in absence of thermo-optic effects. (b) IR transmission spectrum of a fiber taper coupled to a \SI{6.52}{\micro \metre} diameter microdisk. (c) Red line: Visible transmission spectrum of the fiber taper. Green line: SHG signal generated by the IR input light from (b) and collected by the fiber taper. Note that for the $P$ used in (b) and (c), the shortest wavelength IR mode nearly satisfies the double resonance condition.  }
\label{fig:resonances}
\end{center}
\end{figure}

In microdisks,  double resonance is generally not satisfied intrinsically. However, adjusting the microdisk temperature, $T$, can compensate for an initially non-zero relative detuning $|\lambda_c - 2\widetilde\lambda_c|$.  Changes in $T$ modify $\lambda_c$ and $\widetilde{\lambda}_c$ via thermal expansion and the thermo-optic effect \cite{ref:Carmon2004dyn}, and tuning rates $d\lambda_c/dT$ and $d\widetilde{\lambda}_c/dT$ can differ due to dispersion of the modal confinement, refractive index ($n_\text{GaP}$), and thermo-optic coefficient. In the experiment described below, the wavelength dependence of the normalized thermo-optic coefficient $(1/n_\text{GaP})dn_\text{GaP}/dT$ is the dominant mechanism for tuning rate dispersion. For GaP we calculate this coefficient at $\lambda_c$ and $\widetilde{\lambda}_c$ to be $3.4 \times 10^{-5}\ [\text{K}^{-1}]$ and $2.9 \times 10^{-5}\ [\text{K}^{-1}]$ respectively \cite{ref:yas1969opt}, which leads to a differential tuning coefficient $\eta = 1 - (2d\widetilde\lambda_c/dT)/(d{\lambda}_c/dT) = 0.176$.

In our experiment we do not have independent control of $T$. However, because of linear optical absorption and accompanying heating, $T$ is proportional to $P_d(\Delta)$. For a given $\eta$ and initial $|\lambda_c - 2\widetilde\lambda_c|$, it is possible to find a critical power $P = P_\text{crit}$ where double resonance is achieved. The  principle of this scheme is illustrated in Fig.\ \ref{fig:resonances}(a), where we sketch $\Delta$ and $2\widetilde{\Delta} $ versus $\lambda_o$ for three values of $P$, assuming $\eta  > 0$. As $\lambda_o$ is tuned towards $\lambda_c$ from blue to red, the increase in $P_d(\Delta)$  heats the cavity, causing both $\lambda_c$ and $\widetilde{\lambda}_c$ to shift to longer wavelengths. For $P < P_\text{crit}$ (left panel in  Fig.\ \ref{fig:resonances}(a)) the deviation is relatively small and  double resonance is never realized. When $P = P_\text{crit}$ (center panel in Fig.\ \ref{fig:resonances}(a)), the double resonance condition is satisfied.   For $P > P_\text{crit}$ (right panel in Fig.\ \ref{fig:resonances}(a)), ${\lambda}_c$ is shifted past $2\widetilde\lambda_c$ before $\lambda_o$ reaches $\lambda_c$, and double resonance is never achieved.  A sharp change in $\Delta$ and $\widetilde\Delta$  occurs in all three scenarios when $\lambda_o$ ``catches up" to the thermally shifted $\lambda_c$ \cite{ref:Carmon2004dyn}.

We now show that this tuning scheme can be used to demonstrate SHG at double resonance in GaP microdisks fabricated following Mitchell et al.\cite{ref:Mitchell}. To identify promising microdisks for efficient SHG, devices were characterized using fiber taper mode spectroscopy at IR wavelengths, while the spectrum of SHG produced by the microdisk and collected by the fiber taper was simultaneously measured. Figure \ref{fig:resonances}(b) shows the transmission through the fiber taper when it is positioned in the near field of a \SI{6.52}{\micro \metre} diameter microdisk, measured with $P \sim 0.5\,\text{mW}$ from a tunable IR wavelength laser (New Focus TLB-6700). Evanescent coupling from the fiber taper to high-$Q$ microdisk doublet modes \cite{ref:borselli2006high} results in sharp transmission dips and corresponding large $P_d$. Through comparison with finite difference time domain (MEEP) simulations of the microdisk mode spectrum measured from $1470 - 1570\,\text{nm}$, we identify doublets near 1545 nm and 1554 nm as corresponding to TE-polarized modes with $m$, radial ($p$), and axial ($q$) numbers $\{27,1,1\}$ and $\{19,3,1\}$, respectively. From fits to the doublet lineshape, we measured unloaded $Q$ of $1.1 \times 10^5$ and $2.8 \times 10^4$ for the $p = 1$ and $p = 3$ doublets, respectively. Figure \ref{fig:resonances}(c) shows that SHG is observed when pumping both modes, and is strongest for the $p = 1$ mode.

To study the initial alignment between modes near $\widetilde{\lambda}_c$ and the SHG signal, we measured the  fiber taper transmission using a  visible supercontinuum source and spectrometer detection. Comparing the transmission and SHG spectra in Fig.\ \ref{fig:resonances}(c) reveals that the strongest SHG occurs when exciting a doublet mode with  SHG emission at $\widetilde{\lambda}_o = \lambda_o/2$ that is close to a resonance at $\widetilde\lambda_c \sim 772.2\,\text{nm}$  ($\{m,p,q\}=\{56,3,1\}$, $\widetilde Q \sim 9600$). In contrast, SHG from the other mode in the doublet is weak, as it is off resonance from $\widetilde\lambda_c$.

\begin{figure}[t]
\begin{center}
\epsfig{figure=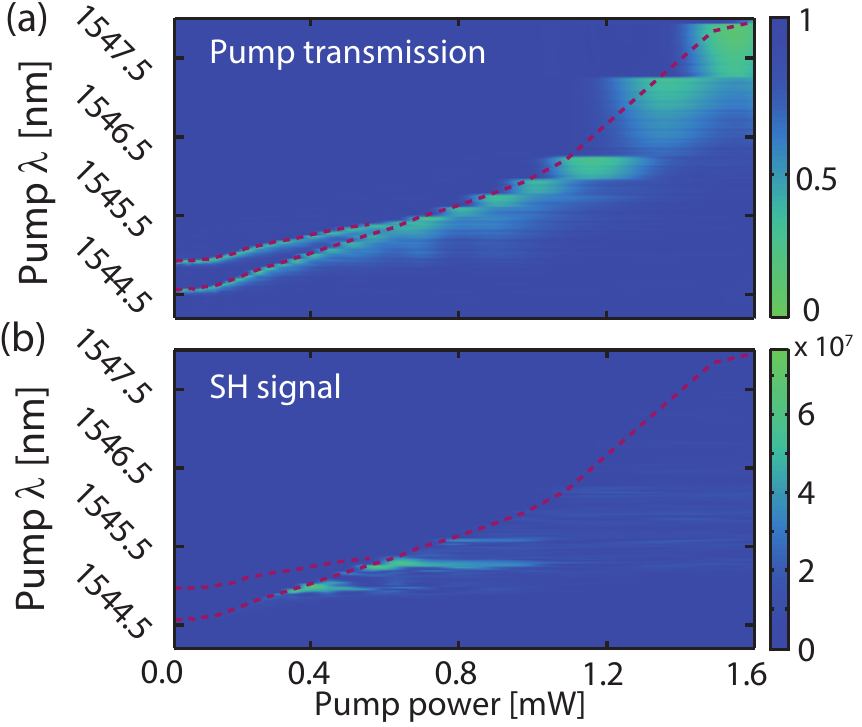, width=1\linewidth}
\caption{
{(a)} Normalized transmission of the pump through the fiber taper when $\lambda_o$ is scanned from 1544 nm to 1548 nm (in the direction of increasing $\lambda_o$) for various fixed values of $P$. A thermally induced cavity resonance shift is evident with increased $P$. The red dashed line traces over points of minimum transmission where $\lambda_o = \lambda_c$. {(b)} SHG signal corresponding to the operating conditions in (a). The  dashed line traces $\lambda_c$ from (a), and serves as a guide to the eye.}
\label{fig:setup}
\end{center}
\end{figure}

Further insight into the role played by double resonance was provided using the tuning scheme described above.  We scanned $\lambda_o$ from blue to red across $\lambda_c$ for a range of fixed values of $P$ and measured the pump transmission (Fig \ref{fig:setup}(a)), and the corresponding SHG (Fig. \ref{fig:setup}(b)) as a function of $P$ and $\lambda_o$. Each point in Figs.\ \ref{fig:setup}(a) and \ref{fig:setup}(b) is the integrated intensity of the narrowband signal detected by the IR and visible spectrometers at $\lambda_o$ and $\widetilde\lambda_o$ respectively, for a given $\{\lambda_o, P\}$. The redshift of $\lambda_c$ in the pump transmission spectrum in Fig.\ \ref{fig:setup}(a) with increasing $P$  results from thermo-optic effects in the microdisk. These effects are also responsible for the  asymmetric ``shark fin'' lineshape evident in Fig.\ \ref{fig:resonances}(b) \cite{ref:Carmon2004dyn}. Note that the jagged edges in Fig.\ \ref{fig:setup}(a) for high $P$  are due to the discrete stepping of $P$. Figure \ref{fig:setup}(b) shows that SHG is most intense when exciting the shorter wavelength mode of the doublet, consistent with the observation in Fig.\ \ref{fig:resonances}(c). Additionally, SHG is maximized at an intermediate value of $P$. As described below, this can be explained by the $P_d$ dependence of $\widetilde\Delta(\lambda_o=\lambda_c)$ illustrated in Fig.\ \ref{fig:resonances}(a). 

The SHG behaviour can be investigated more quantitatively by analyzing the subset of data in Fig.\ \ref{fig:setup} where the pump is on-resonance ($\Delta = 0$).  In this case the SHG signal is described by a simplified form of Eq. \eqref{eq:eff} \cite{ref:Kuo2011,ref:lake2015sup}:
\begin{equation}
\widetilde{P}(\Delta = 0)\propto
\frac{(\widetilde{\lambda}_c/2\widetilde{Q}_t)^2}{\widetilde{\Delta}(\lambda_o,P)^2+(\widetilde{\lambda}_c/2\widetilde{Q}_t)^2} \times |K|^2 P^2
\label{eq:poweff}
\end{equation}
This expression is a product of two functions: the first captures the impact of the $P$ dependence of the detuning $\widetilde\Delta$ between the SHG signal ($\widetilde\lambda_o = \lambda_c/2$) and the nearest cavity mode ($\widetilde\lambda_c$), and the second describes the usual $P^2$ dependence of SHG, and scales with the effective nonlinear susceptibility of the microdisk modes of interest. Conveniently, $\widetilde{\Delta}$ is a linear function of $P$ and $\lambda_o$ for $\Delta = 0$ (see Supplementary Information \cite{ref:lake2015sup}), allowing Eq.\ \eqref{eq:poweff} to be easily fit to experimental data.

\begin{figure}[t]
\begin{center}
\epsfig{figure=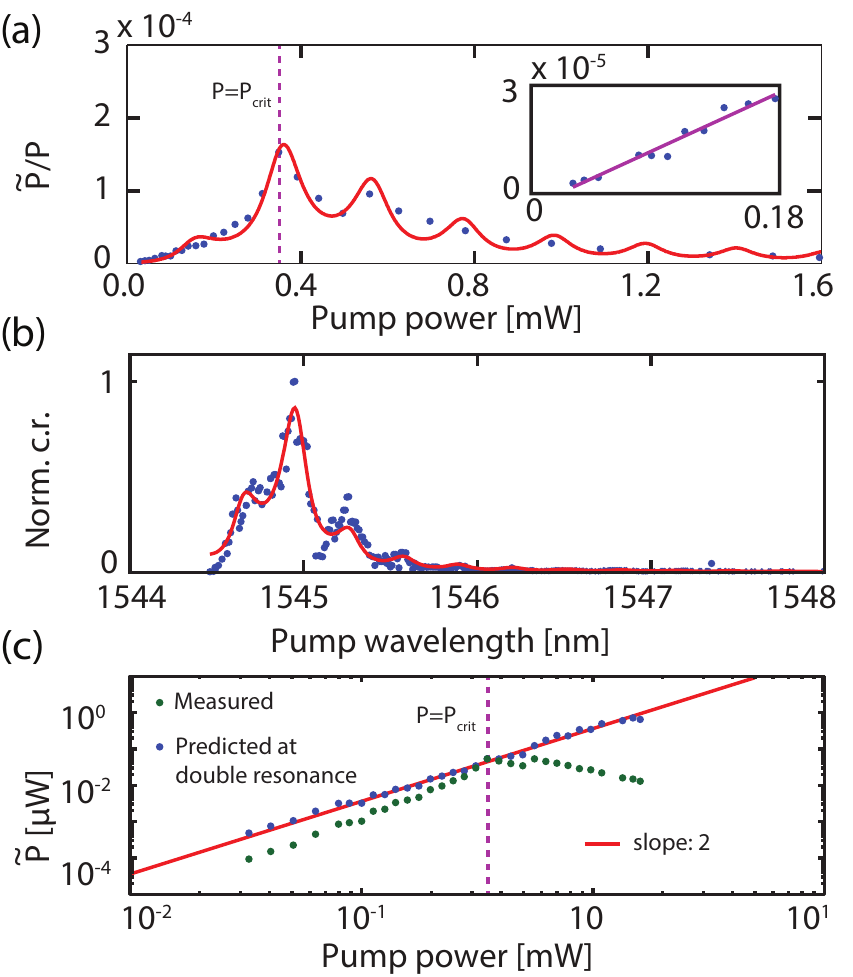, width=1\linewidth}
\caption{(a)  SHG absolute efficiency $\widetilde P/P$ versus $P$ for $\Delta = 0$. The red line is a fit to the data using Eq.\ \eqref{eq:poweff}. The inset highlights low $P$ data and includes a linear fit. (b) Normalized SHG count rate ($\propto \widetilde P/P^2$) plotted versus pump wavelength $\lambda_o$ for $\Delta = 0$, where the red line is a fit to the data adapted from Eq.\ \eqref{eq:poweff}. (c) SH power  vs.\ $P$ for $\Delta=0$. The green points show the unprocessed measured data. The blue points show the predicted SH power if $\widetilde\Delta =0$ for the entire range of $P$. The red line is a weighted least squares fit to the data of a function proportional to $P^2$, and indicates a normalized SHG efficiency $\widetilde P/P^2 = 3.8\pm 0.2 \times 10^{-4} \,\text{mW}^{-1}$.}
\label{fig:eff}
\end{center}
\end{figure}

In Fig.\ \ref{fig:eff} we analyze the $\Delta=0$ data from Fig.\ \ref{fig:setup} using Eq.\ \eqref{eq:poweff}. We only consider the shorter $\lambda_c$ doublet mode, and find $\lambda_c(P;\Delta = 0)$ from the data in Fig.\ \ref{fig:setup}(a). In Fig.\ \ref{fig:eff}(a), we plot SHG absolute efficiency vs.\ $P$ for $\Delta=0$. Here absolute efficiency is defined as $\widetilde P/P$, where powers are measured in the fiber taper immediately before ($P$) and after ($\widetilde P$) the microdisk, taking into account asymmetric fiber taper insertion loss.  We emphasize that because of the $P$ dependence of $\lambda_c$ shown in Fig.\ \ref{fig:setup}(b), for every value of $P$, the selected $\lambda_o$ corresponding to  $\Delta =0$ varies. For low $P < 0.18$ mW, the absolute efficiency is observed to increase approximately linearly with $P$, as thermo-optic effects are small compared to the intrinsic ``cold cavity'' $\widetilde\Delta|_{P=0}$. In this region we estimate $\widetilde\Delta$ to range from $-34\,\text{pm}$ to $-28\,\text{pm}$.  As $P$ approaches $P = P_\text{crit} = 0.35\,\text{mW}$,  thermo-optic effects become significant, $\widetilde\Delta \to 0$ satisfying double resonance, and the absolute efficiency increases superlinearly with $P$ to a maximum value of $1.5\times10^{-4}$ . For $P>P_\text{crit}$, absolute efficiency is observed to decrease. 

The observed $P$ dependence of $\widetilde P(\Delta=0)$ is consistent with $\widetilde\Delta(P;\Delta=0)$ increasing monotonically with $P$ from $\widetilde\Delta < 0$ to $\widetilde\Delta > 0$. We test this quantitatively by comparing the predictions from Eq.\ \eqref{eq:poweff} with the measured data in Fig.\ \ref{fig:eff}(a). Here the required  fitting parameters are an overall scaling factor, and the differential thermo-optic tuning coefficient $\eta$ that determines the linear dependence of $\widetilde\Delta$ on $P$ for $\Delta = 0$.  We find a best fit for $\eta = 0.181$, which is within $3\%$ of the theoretical value presented above. In applying this model, we included a $\lambda_o$ dependence in the scaling factor that follows an Airy function and accounts for etaloning in the non-wedged neutral density filters used in this experiment, resulting in  oscillations evident in the data. 

Further analysis of the SHG signal at $\Delta=0$ is given in Fig.\ \ref{fig:eff}(b), which shows the SHG signal for varying $\lambda_o$.  Here we are plotting $\widetilde P$ normalized by $P^2$ in order to isolate the cavity contribution described in Eq.\ \eqref{eq:poweff}.  The resulting fit of Eq.\ \eqref{eq:poweff} to the data, shown in Fig.\ \ref{fig:eff}(b), has good agreement.  For this fit, we use the value for $\eta$ found above, and treat  $Q_t$ as a fitting parameter; the fit predicts $\widetilde Q_t \sim 1.0\times10^4$, in close agreement with the spectrometer measured value in Fig.\ \ref{fig:resonances}(c). As in the $P$-dependent fit in Fig.\ \ref{fig:eff}(a), we included a $\lambda_o$ dependent scaling factor to account for etaloning from the neutral density filters that creates the observed oscillations.

Finally, in Fig.\ \ref{fig:eff}(c) we analyze the SHG signal power vs.\ $P$ at $\Delta = 0$. In this analysis, we plot $\widetilde P$ in absolute measured units, as well as scaled by the $P$ dependent microdisk resonance response described by the first term in Eq.\ \eqref{eq:poweff} and predicted from the fit to $\widetilde\Delta(P;\Delta=0)$ obtained for Fig.\ \ref{fig:eff}(a). This latter scaling isolates the term in Eq.\ \eqref{eq:poweff} that describes the microdisk nonlinear susceptibility, and is ideally proportional to $|K|^2 P^2$.   As shown in Fig.\ \ref{fig:eff}(c), when scaled in this manner, the data is well described by a  $P^2$ dependence. This plot represents the predicted SHG signal if the enhancement provided by the microdisk density of states is fixed to its maximum value at $\widetilde\Delta = \Delta = 0$. Note that in our experiment, this efficiency is only realized when $P=P_\text{crit}$, as evident by comparing the unscaled and predicted data in Fig.\ \ref{fig:eff}(c).

The maximum normalized outside efficiency, defined as $\widetilde{P}/P^2$, was found from the point of highest efficiency in Fig.\ \ref{fig:eff}(a) to be $4.4\times10^{-4}$ \SI{}{mW}$^{-1}$ for the case $P=P_\text{crit}$. The weighted least squares fit to the scaled data shown in Fig.\ \ref{fig:eff}(c) can be used as a consistency check, and gives the predicted normalized outside efficiency if the double resonance condition $\Delta = \widetilde\Delta = 0$ is satisfied over the entire range of $P$.  We find a normalized efficiency of $(3.8 \pm 0.2)\times10^{-4}$ \SI{}{mW}$^{-1}$, in good agreement with the measured maximum normalized outside efficiency at $P_\text{crit}$  in Fig.\ \ref{fig:eff}(b) where the double resonance condition is satisfied. In comparison, the SHG signal at low $P$ shown in the inset Fig.\ \ref{fig:eff}(b), where $\widetilde\Delta$ is approximately constant (i.e.\ independent of $P$),  has a normalized outside efficiency of $1.78\times10^{-4}$ \SI{}{mW}$^{-1}$. This illustrates that the SHG normalized outside efficiency $\widetilde P/P^2$, in addition to the total efficiency $\widetilde P/P$ shown in Fig.\ \ref{fig:eff}(b), is enhanced through satisfaction of the double resonance condition at $P = P_\text{crit}$.

In conclusion, we have demonstrated resonant second harmonic generation from 1550 nm to 775 nm in a GaP microdisk with maximum normalized outside efficiency of $3.8\times10^{-4}$ \SI{}{mW}$^{-1}$ for $P = P_\text{crit} = 0.35\,\text{mW}$, which is larger than previously reported values in similarly sized structures used for 2000 nm to 1000 nm wavelength conversion \cite{ref:Kuo2014}. We have shown that this efficiency is achieved via double resonance between high-$Q$ modes of the GaP microdisk at both IR ($Q \sim 1.1 \times 10^5$) and visible ($\widetilde Q \sim 1.0 \times 10^4$) wavelengths. Further improvements to the efficiency could be made by spectrally aligning higher-$\widetilde Q$ modes, and optimizing fiber taper coupling at visible wavelengths. Control of the differential tuning independent of pump power could be realized by thermally heating using light from an additional optical mode not involved in the nonlinear conversion process. As GaP is a piezoelectric material, it may be possible to tune the modes through electronic means \cite{ref:wong2004str}. The geometry presented in this letter can also be adapted for use in other nonlinear optics scenarios such as four wave mixing and downconversion \cite{ref:Li2015chi, ref:Jiang2015sil, ref:yang2007gen, ref:helt2010spo}. With the present system we have already observed third harmonic generation, although efficient collection of the resultant \SI{515}{\nano \meter} light via fiber taper requires additional optimization.

\bibliography{SHG_bib}

\newpage

\setcounter{equation}{0}
\setcounter{figure}{0}
\setcounter{table}{0}
\setcounter{page}{1}
\makeatletter
\renewcommand{\theequation}{S\arabic{equation}}
\renewcommand{\thefigure}{S\arabic{figure}}

\section{Thermal tuning}

The thermal dependance of the laser cavity detuning can be approximated by the expression \cite{ref:Carmon2004dyn}:
\begin{equation}
\Delta(\delta T) \approx \lambda_o-\lambda_c \left [1+a \delta T \right],
\label{eq:Lambdashift}
\end{equation}
with $a=\epsilon+ \frac{1}{n}\frac{dn}{dT}$. Here $\delta T$ is the deviance from the equilibrium temperature, $\epsilon$ is the thermal expansion coefficient, and $n$ is the refractive index. 

Furthermore, we can write an expression for the change in cavity temperature due to optical absorption when the cavity is pumped at resonance and is in thermal equilibrium:
\begin{equation}
\Delta T = \frac{2Q^2}{Q_\kappa Q_{abs}}\frac{P_f}{K_{th}}=c_{th}P_f,
\label{eq:Tshift}
\end{equation}
where $Q_{abs}$ is the quality factor associated with optical absorption in the cavity, and $K_{th}$ is the inverse of the thermal time constant of the cavity. This expression is derived by considering the rate at which power is absorbed into the cavity when optically pumped on resonance, and equating it to $K_{th}\delta T$, the rate at which power is dissipated. For simplicity we have grouped together terms into the constant $c_{th}$

Making use of equations \eqref{eq:Tshift} and \eqref{eq:Lambdashift} we can write expressions for $\widetilde{\Delta}$ for the case when $\Delta=0$:
\begin{equation}
\widetilde{\Delta} = 
\begin{cases} 
a_1 \lambda_o+a_0 \\
b_1 P_f + b_0,
\end{cases}
\label{eq:para}
\end{equation}
where:
\begin{align}
a_0&= -\left[1 - \frac{\widetilde{a}}{a} \right]\widetilde{\lambda}_o,\\ 
a_1&= \frac{1}{2}-\frac{\widetilde{a}\widetilde{\lambda}_o}{a\lambda_o},\\
b_0&=c_{th}\left[ \frac{a\lambda_0}{2} - \widetilde{a}\widetilde{\lambda}_o\right],\\
b_1&=\frac{\lambda_o}{2}-\widetilde{\lambda}_o.
\end{align}

\section{SHG in doublets}

\begin{figure*}[t]
\begin{center}
\epsfig{figure=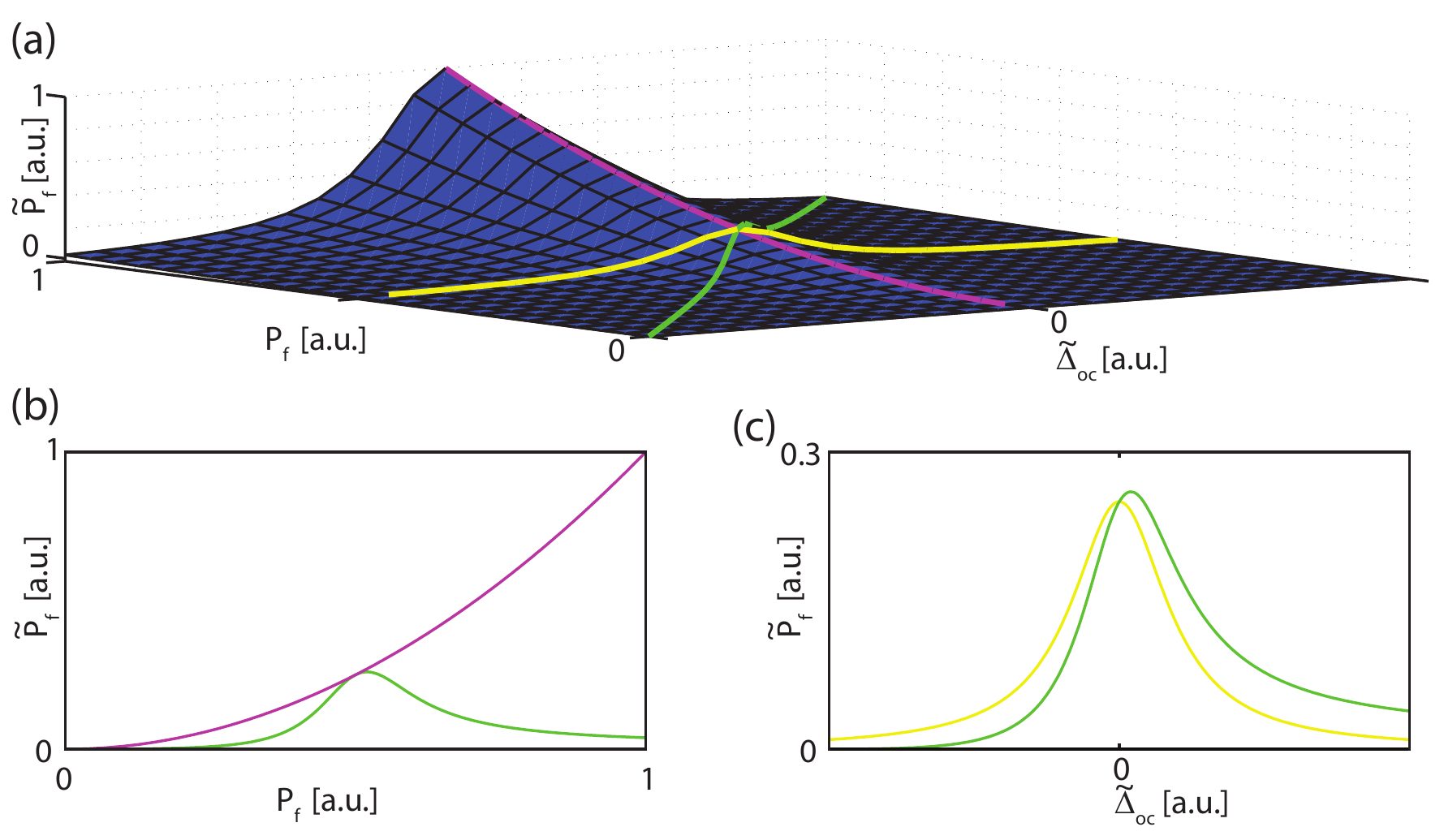, width=1\linewidth}
\caption{(a) Second harmonic signal for the special case $\Delta=0$, as a function of $P_f$ and $\widetilde{\Delta}$. The coloured curves represent possible trajectories of an experiment. (b) Power dependence of second harmonic generation for the case of constant detuning (magenta curve) and for variable detuning (green curve). (c) Detuning dependence of second harmonic generation of constant power (yellow curve), and for varied power (green curve).}
\label{fig:eqfig}
\end{center}
\end{figure*}

The optical resonance used in the current experiment was a doublet mode. Such modes occur in whispering-gallery type resonators when surface roughness causes the coupling rate between clockwise and counter clockwise propagating mode to exceed the combined loss rate to all other channels \cite{ref:borselli2006high}. The transmission spectra of the doublet contains two distinct resonances, each of which is a standing wave created from linear combinations of clockwise and counter clockwise propagating modes. In the case of the doublet considered in the current experiment, these standing waves are spectrally well separated, which permits the individual excitation of each of these modes. We write the amplitude of the standing wave at the fundamental frequency $\omega_0$, as\cite{ref:borselli2006high}: 
\begin{equation} \label{eq:ampfun}
A=\frac{-\kappa/\sqrt{2}S_f}{\imath(\Delta_{\omega})-\gamma_t/2},\\
\end{equation}
where $A$ is the amplitude of the standing wave mode, $S_f$ is the amplitude of the forward propagating mode in the fiber taper, $\kappa$ is the waveguide-cavity coupling, $\Delta_{\omega}$ is the laser-cavity detuning in terms of frequency, and $\gamma_t$ is the energy decay rate of the cavity into all channels. Here $|A|^2$ is normalized to energy, and $|S_f|^2$ is normalized to power. 

In contrast to the travelling wave fundamental modes considered elsewhere \cite{ref:Kuo2011}, a standing wave fundamental mode will excite both clockwise and counter clockwise modes at the second harmonic frequency $\widetilde{\omega}_o=2\omega_o$. The coupled mode equations describing this process may be written as:
\begin{align}
\frac{d\,\widetilde{a}^{ cw }}{dt} &= (\imath\widetilde{\omega}_c-\widetilde{\gamma}_t/2)\widetilde{a}^{ cw}+\widetilde{\zeta}, \label{eq:cmEqn1}\\ 
\frac{d\,\widetilde{a}^{ccw}}{dt} &= (\imath\widetilde{\omega}_c-\widetilde{\gamma}_t/2)\widetilde{a}^{ccw}+\widetilde{\zeta},\label{eq:cmEqn2}
\end{align}
where $\zeta$ is the second harmonic source term, $\widetilde{\gamma}_t$ is the energy decay rate of the loaded cavity around the cavity frequency $\widetilde{\omega}_c$, and $\widetilde{a}^{cw,ccw}=\widetilde{A}^{cw,ccw}e^{\imath\omega_o t}$ with $\omega_o$ representing the laser frequency and $\widetilde{A}^{cw,ccw}$ as the amplitude of the clockwise or counterclockwise propagating modes. Note that in our experiment, the second harmonic mode was found to be a singlet mode, so surface roughness did not appeciably couple the clockwise and counter clockwise propagating modes at $\widetilde{\omega}_c$.

The explicit form of the nonlinear source term may be written as:
\begin{equation}\label{eq:sourceshunsimplified}
\widetilde{\zeta}= 2\pi \frac{\widetilde{v}_g}{2 \pi R} \left| \left\langle \frac{d \widetilde{A}^{cw,ccw}}{d \theta} \right\rangle \right|
\end{equation}
where $\widetilde{v}_g$ is the group velocity of the SH mode, and $R$ is the radius of the disk. The first term of this expression gives the frequency at which the SH mode circulates the disk in units of rad/s, and the second term describes the average gain per cycle. 

The nonlinear source term can be calculated from the spatial profile of the optical modes as:
\begin{equation}\label{eq:sourcesh}
\widetilde{\zeta}=\frac{A^2\delta\omega_{FSR}}{4 \pi}\int_0^{2\pi} \left[K_+e^{\imath (\Delta m +2)\theta}+K_-e^{-\imath (\Delta m +2)\theta} \right] d\theta
\end{equation}
where $\widetilde\zeta$ is the source term, $\delta\omega_{FSR}$ is the free spectral range of the microdisk at $\omega_0$, and $K_+$, $K_-$ are the second harmonic coefficients, and $\Delta m$ is the mismatch in the azimuthal numbers of the microdisk at $\omega_0$ and $\widetilde{\omega}_0$, defined as $\Delta m \equiv \widetilde{m}-2m$. The explicit form of the second harmonic coefficients, as derived in \cite{ref:Kuo2011}, may be written as: 
\begin{align} 
K_+&=-\frac{d_{14}}{2\epsilon_0\widetilde{\omega}_0n^4}\int_0^\infty \int_{-h/2}^{h/2} \rho \widetilde{\psi} \left( \frac{m\psi}{r} + \frac{\partial\psi}{\partial \rho} \right)^2 dz d\rho, \\ 
K_- &= \frac{d_{14}}{2\epsilon_0\widetilde{\omega}_0n^4}\int_0^\infty \int_{-h/2}^{h/2} \rho \widetilde{\psi} \left( \frac{m\psi}{r} - \frac{\partial\psi}{\partial \rho} \right)^2 dz d\rho, 
\end{align}
where $n$ is the refractive index of GaP at $\omega_0$, and $\psi$, $\widetilde{\psi}$ give the radial and vertical dependence of the $\hat{z}$-component of the modes at $\omega_0$ and $\widetilde{\omega}_0$. Here we have assumed the normalization conditions $\int |\psi|^2 dA=\delta\omega_{fsr}/2\pi$ and $\int |\widetilde{\psi}|^2 dA=\delta\widetilde{\omega}_{fsr}/2\pi$ where the integrals are taken over the infinite half-plane \cite{ref:Kuo2011}.

Finally we use \ref{eq:cmEqn1}, \ref{eq:cmEqn2} to solve for the steady state amplitudes of the second harmonic modes propagating in the clockwise and counter clockwise directions:
\begin{align}\label{eq:ampsh}
\widetilde{A}^{ cw} &= \frac{\widetilde{\zeta}}{\imath \widetilde{\Delta}_{\omega}-\widetilde{\gamma}_t/2},\\ 
\widetilde{A}^{ccw} &= \frac{\widetilde{\zeta}}{\imath \widetilde{\Delta}_{\omega}-\widetilde{\gamma}_t/2},
\end{align}
where $\widetilde{\Delta}_{\omega}$ is the laser-cavity detuning $\omega_o/2-\widetilde{\omega}_c$. 

Making use of the amplitude normalizations, and waveguide-cavity coupling parameters, we can relate ${\widetilde{P}_{f,b}}$, the power in the taper at $\widetilde{\lambda}_o$ in either the forward or backward direction and $P_f$, the power in the taper at $\lambda_o$ to the amplitudes:
\begin{align}
P_f &= |S_f|^2,\label{eq:powIn}\\ 
\widetilde{P}_{f,b} &= |\widetilde{S}_{f,b}|^2=\frac{|\widetilde{\kappa}|^2|\widetilde{A}_{f,b}|^2}{2}\label{eq:powOut}
\end{align}

Combining equations (\ref{eq:ampfun}), (\ref{eq:sourcesh}), (\ref{eq:ampsh}), (\ref{eq:powIn}), (\ref{eq:powOut}) and writing the decay rates in terms of their respective quality factors we arrive at a final expression in terms of $\widetilde{P}_{f,b}$ and $P_f$:
\begin{multline}
\widetilde{P}_{f,b}=|K|^2 
\frac{\widetilde{\lambda}_c/2\widetilde{Q}_\kappa}{\widetilde{\Delta}^2+(\widetilde{\lambda}_c/2\widetilde{Q}_t)^2}
\left[ \frac{1}{2}\frac{\lambda_c/2Q_\kappa}{\Delta^2+(\lambda_c/2Q_t)^2} \right]^2
 P_f^2 .
\label{eq:Full}
\end{multline}
with:
\begin{multline}
|K|^2=\frac{\lambda_c}{2\pi c}\delta \lambda_{FSR}^2 \big [ K_+e^{\imath\pi(\Delta m+2)}\text{sinc} [\pi(\Delta m+2 )] \\
+K_-e^{\imath\pi(\Delta m-2)}\text{sinc} [\pi(\Delta m-2 )]\big]^2.
\end{multline}

In order to better visualize the physics behind the data analysis of the main text, we have plotted Eq. \eqref{eq:Full} in Fig. \ref{fig:eqfig} for the case $\Delta=0$. In this instance, the expression for $\widetilde{P}_f$ is seperable into a function of $P_f$ and a function of $\widetilde{\Delta}$. We display this on Fig. \ref{fig:eqfig} along with three curves representing various paths in $P_f-\widetilde{\Delta}$ space which an experiment might take. The magenta curve represents the case of fixed detuning $\widetilde{\Delta}=0$, which demonstrates the typical quadratic dependence on $P_f$ expected from a second harmonic experiment. The yellow curve is the case of fixed power, with a variable detuning, which displays the Lorentzian line shape contributed by the cavity resonance lineshape. Lastly, the green curve represents the case where both detuning and pump power are varied.

We would like to thank Aaron C.\ Hryciw, and J.P.\ Hadden for their assistance with this project. This work was supported by NRC, CFI, iCORE/AITF, NSERC, and CNPq.

\bibliography{SHG_bib}

\newpage
\newpage

\end{document}